\documentclass[twocolumn]{svjour3}
\usepackage{amsmath,amsfonts,amssymb}
\usepackage{graphicx}
\usepackage{setspace} 
\usepackage{tocloft}
\usepackage{textcomp}
\usepackage{tocloft}
\usepackage[flushleft]{threeparttable}
\usepackage{natbib}
\usepackage{lipsum}
\usepackage{float}
\usepackage{amsmath}
\usepackage{latexsym}
\usepackage{booktabs,etoolbox}
\usepackage[utf8x]{inputenc}
\usepackage{tablefootnote}

\newcommand{\as}{$^{\prime\prime}$}

\newcommand{\aap}{    {\it Astron. Astrophys.}}
\newcommand{\aas}{    {\it Astron. Astrophys.}}

\newcommand{\aj}{     {\it Astron. J.}} 
\newcommand{\ea}{     {\it Exp. Astron.}}

\newcommand{\apj}{    {\it Astrophys. J.}}
\newcommand{\apjs}{   {\it Astrophys. J. Suppl.}}
\newcommand{\apjl}{   {\it Astrophys. J. Let.}}
\newcommand{\apss}{   {\it Astrophys. Space Sci.}}

\newcommand{\mnras}{  {\it Mon. Not. R. Astron. Soc.}}

\newcommand{\pasp}{   {\it Pub. Astron. Soc. Pac.}}

\newcommand{\ssr}{    {\it Space Sci. Rev.,}}

\newcommand{\jatis}{  {\it J. Astron. Tel. Instr. Sys.}}
\newcommand{\jai}{    {\it J. of Astron. Instr.}}
\newcommand{\procspie}{{\it Proc. SPIE}}

\begin{document}

\title{The Near Ultraviolet Transient Surveyor (NUTS): An ultraviolet telescope to observe variable sources}

\author{S.~Ambily \and Mayuresh Sarpotdar \and Joice Mathew \and Binukumar G.~Nair \and A.~G.~Sreejith \and Nirmal K.~ \and Jayant Murthy \and Margarita Safonova \and Rekhesh Mohan \and Vinod Kumar Aggarval \and S.~Nagabhushanam \and Sachin Jeeragal}

\institute{S.~Ambily \at
Indian Institute of Astrophysics, Koramangala 2nd block, Bangalore, 560034, India \\
Laboratory for Atmospheric and Space Physics, University of Colorado, UCB 600, Boulder, CO, 80309, USA \\
\email{ambily.suresh@colorado.edu} \and
Mayuresh Sarpotdar \at Satellite Research Centre, Nanyang Technological University, Singapore \and
Joice Mathew \at Advanced Instrumentation and Technology Centre, RSAA, Mount Stromlo Observatory, Australian National University, Canberra, Australia \and
Binukumar Gopalakrishnan Nair \and Jayant Murthy \and Margarita Safonova \and Rekhesh Mohan \at Indian Institute of Astrophysics, Koramangala 2nd block, Bangalore, 560034, India \and
A.~G.~Sreejith \at Space Research Institute, Austrian Academy of Sciences, Schmiedlstrasse 6, Graz, Austria \\
Laboratory for Atmospheric and Space Physics, University of Colorado, UCB 600, Boulder, CO, 80309, USA
\and
K.~Nirmal \at Deutsches Elektronen-Synchrotron (DESY), Zeuthen, Germany
\and 
Vinod Kumar Aggarval \and S.~Nagabhushanam \and Sachin Jeeragal \at
Crucible of Research and Innovation, PES Institute of Technology, Bangalore, India}

\date{Received: date / Accepted: date}

\maketitle



\begin{abstract}
Observing the ultraviolet (UV) sky for time-varyiable phenomena is one of the many exciting science goals that can be achieved by a relatively small aperture telescope in space. The Near Ultraviolet Transient Surveyor (NUTS) is a wide-field ($3^\circ$) imager with a photon-counting detector in the near-UV (NUV, 200~\textendash~300~nm), to be flown on an upcoming small satellite mission. It has a Ritchey–Chrétien (RC) telescope design with correction optics to enable wide-field observations while minimizing optical aberrations. We have used an intensified CMOS detector with a solar blind photocathode, to be operated in photon-counting mode. The main science goal of the instrument is the observation of transient sources in the UV, including flare stars, supernovae, and active galactic nuclei. NUTS's aperture size and effective area enable observation of relatively unexplored, brighter parts of the UV sky which are usually not accessible to larger missions. We have designed, fabricated, and assembled the instrument, and the final calibrations and environmental tests are being carried out. In this paper, we provide the scientific motivation and technical overview of the instrument and describe the assembly and calibration steps. 
\end{abstract}


\keywords{Variable sources \and Ultraviolet astronomy \and Small satellites \and Wide field imager \and UV sky surveys} 

\section{Introduction}

CubeSats and small satellites present a unique opportunity for space-based astronomy by providing a platform for technology development while exploring specific science goals that are difficult to achieve with large missions. This is especially true for the ultraviolet (UV) wavelength range where there are few missions that are currently active: the UV spectrographs on the {\it Hubble Space Telescope} \citep{stis_woodgate_1998, cos_green_2012}; the Ultraviolet Imaging Telescope (UVIT)\footnote{Currently (2021) only the FUV channel is operational.} on {\it AstroSat} \citep{Tandon_UVIT_2017}; and the supplementary UV telescopes on the {\it Swift} \citep{uvot_roming_2005}; and {\it XMM-Newton} \citep{Mason_xmm_2001} missions. The last few years have seen an increased interest in the UV region\footnote{The UV spectrum is usually subdivided into a number of ranges (by the ISO standard ISO-21348): Vacuum UV: $10-200$ nm; Extreme UV: $10-121$ nm; Far UV: $121-200$; and Near UV: $200-400$ nm.} with a number of CubeSat missions under development \citep{cute_fleming_2018, sparcs_shkolnik_2018, Fleming_sprite_2019, Cenko_Dorado_2019}, each of which seeks to address a set of distinct science goals. CubeSats also provide a platform for technological advancements for the future that were traditionally carried out using relatively low-cost sounding rockets and high-altitude balloon experiments \citep{france_rockets_2016, Sreejith_balloon_2016}. There are several technological challenges to be addressed for future UV missions, such as improving the performance of the detectors in the Far-UV (FUV), developing high-reflectivity mirror coatings for the vacuum UV, and fabricating low-noise large-area microchannel plates \citep{luvoir_report_2019, habex_report_2020}. CubeSats offer an excellent opportunity to improve the technology readiness levels of these activities by providing actual spaceflight opportunities at faster time scales and a fraction of the cost of full-fledged satellite missions. 

Space missions such as {\it GALEX} \citep{Galex_martin_2005}, {\it FUSE} \citep{fuse_moos_2000}, and the suite of UV instruments on the {\it HST} have demonstrated the fascinating science that can be done with UV observations. Although these missions have studied the sky in both FUV and NUV through both photometric and spectroscopic observations, many areas in the UV sky are not yet well-studied. For example, the {\it GALEX} mission has imaged around 80\% of the sky, with a limiting magnitude of 28 AB \citep{bianchi_GalCat_2017}. However, due to detector limitations \citep{galex_morrisey_2007}, it has not been able to observe the brightest and most interesting regions of the sky such as the Milky Way plane and the Magellanic Clouds. A small dedicated telescope can survey those areas that are typically left out by large missions with broader science motivations \citep{Noah_Cubesat_2014}. 

We, the small-payloads group at the Indian Institute of Astrophysics (IIA), have been working on building and testing payloads targeted at astronomical and atmospheric observations in the UV \citep{Mathew_2017, bharat_sing_2020}. We have used high-altitude balloons to test the sub-system prototypes and later develop them into full-fledged space payloads \citep{Sreejith_balloon_2016, Mathew_2018}. The Near Ultraviolet Transient Surveyor ({\it NUTS}) is one such payload and is an NUV imaging telescope that we plan to fly on a small satellite. The instrument is a 150-mm Ritchey–Chrétien (RC) telescope covering the wavelength range of 200~\textendash~300~nm using a solar-blind photocathode. The instrument has a peak effective area of 12~cm\textsuperscript{2} and can detect objects as faint as 21 AB \citep{OkeGunn_AB}. With a wide field of view (FOV: $3^\circ$), we will search for variable UV sources such as flare stars, supernovae, and other transient events with a spatial resolution of $13^{\prime\prime}$. These observations will be used to generate calibrated light curves of interesting transient phenomena with a cadence of up to 2~ms. Because the aperture and the effective area of the telescope are small, our primary targets are bright regions of the UV sky where larger missions are not able to observe due to detector limitations.

\begin{table*}[ht!]
\begin{center}
\begin{tabular}{lcccc}
\hline
Mission parameter & {\it GALEX}  & {\it UVIT NUV}  & {\it ULTRASAT}  & {\it NUTS} \\
& \citet{galex_morrisey_2007} &\citet{Tandon_UVIT_2017} & {\it weizmann.ac.il/ultrasat/} & This work \\\hline
Aperture (mm)           & 500          &  376     &  330     &  146   \\
Collecting area~(cm$^2$) & 1950         & \bf 852   & 856      & 141 \\ 
Detector size           &4K$\times$4K & 512$\times$512 & 4 $\times$ 4K$\times$4K & 1K$\times 1$K\\  
Pixel scale ($^{\prime\prime}$/pix) & 1.5           &  3       & 5.4      &  4.3     \\
Spatial resolution ($^{\prime\prime}$)  &  5.3         & $1.2-1.6$& 13      &  13      \\
FOV (deg$^2$)           & 1.21         & 0.18     & 200     &  7.07    \\
Wavelength range (nm)              & $177-283$    &$200-300$ & $220-280$& $200-300$\\   
Brightness limit (AB)&  9          &  9       &  16       & 7        \\
Weight (kg)             &    280       &    230   &  400   & $\sim$4.5\\ 
Approximate cost (\$$)$
        & 150M     & 6M    & 100M  & 0.2M \\
Current status          & Decommissioned &  Non-functional   & In prep. & In prep. \\
\hline
\end{tabular}
\end{center}
\caption{Comparison of missions parameters in NUV channel. The cost figures are taken from official releases, and may not include the launch and operation costs in some cases. }
\label{table:parameters}
\end{table*}

\section{Scientific Motivation}

Despite its scientific potential, time domain astronomy in the UV has not been studied extensively \citep{ultrasat_sagiv_2014, Mathew_2017}. Traditionally, transient observations from space are carried out in the high energy domains of X-Rays and gamma rays. In addition, ground-based surveys such as the Palomar \citep{PTF_law_2009} and Zwicky \citep{ZTF_bellm_2019} Transient Facilities, and the Vera C.~Rubin Observatory Legacy Survey of Space and Time ({\it LSST}, \cite{lsst_2019}) are also capable of covering large swaths of the sky to look for variable sources. It is important to complement these with space-based UV observations, as in many cases the UV transient can be observed ahead of time and can act as an alert for observations from the ground \citep{ultrasat_sagiv_2014}. The UV sky background is faint, making a majority of variable sources in the night sky detectable with a small space-based observatory \citep{safonova_luci_2014}. The potential of a wide-field UV survey for transients detection and observations has been demonstrated by the {\it GALEX} time domain survey that discovered and classified over a thousand variable UV sources \citep{galex_gezari_2013}. However, although UV space missions of the past and present, such as {\it GALEX} or {\it SWIFT}, have resulted in the discovery of many interesting UV transients, a dedicated program is not feasible with such shared resources but is ideal for a small satellite mission. 

A comprehensive overview of the transient astronomical sources that can be observed in the UV range can be found in \cite{ultrasat_sagiv_2014} and \cite{wang_2019}. We have also outlined a number of scientific objectives of our balloon program in \cite{Mathew_2018}. A dedicated small imaging payload can survey the night sky for bright transient sources such as supernovae \citep{welsh_SN_2011, Wils_novae_2010, ganot_SN_2016}, Active Galactic Nuclei \citep{Wang_cl-agn_2019}, flare stars \citep{Welsh_mdwarf_2007}, and gravitational wave counterparts \citep{gluv_harper_2017}. Surveying the UV sky for such transients, along with the follow-up observations in ground-based optical and radio wavelengths, are critical steps that can help answer key astrophysical questions. 

Some of the most exciting transients are supernova explosions which arise from the death of massive stars and are associated with a prominent high-energy emission. Observations of supernovae play a vital role in understanding the star formation rates and chemical evolution of the universe \citep{Rabinak_2011}. Determining the physical properties of massive stars prior to the explosion is a critical step towards understanding and constraining the current models of stellar evolution. The radiation from early stages of exploding supernovae called shock-breakout flares, as well as the following cooling phase, are observable in the UV wavelengths \citep{ganot_SN_2016, SN_UV_NakarSari_2010}. A modest NUV camera can contribute greatly to the early detection and observation of supernovae explosions \citep{ultrasat_sagiv_2014}. Observing these UV flare parameters such as duration and amplitude helps in estimating the pre-explosion conditions of their progenitor stars such as the radius and chemical composition \citep{Rabinak_2011}.

Stellar flares are another poorly understood astronomical transients that can greatly benefit from UV observations. Late-type stars such as K and M dwarfs, having a high amount of activity in their corona and strong magnetic fields associated with them, flare quite frequently in the UV with a $10^2$ to $10^3$ times increase in intensity \citep{Welsh_mdwarf_2007, Brasseur_flares_2019}. Such stars are interesting candidates for exoplanet science, as many of these stars host close-in exoplanets making them viable for transit observations \citep{Howard_exoplanets_2012, dressing_2015}. Detection and characterization of terrestrial planets are limited to cool stars for the next few years due to observational constraints. But to characterize planet properties through transit observations, it is critical to understand the stellar activity variations and separate those from planetary signals. 

M dwarf flares, especially in the far and extreme UV, are also important from a habitability perspective, as the high energy associated with these flares can impact the planetary atmospheric composition and thereby the habitability \citep{Rugheimer_UVrad_2015, johnstone_hab_2015}. It is also proposed that the flares are associated with an intense NUV radiation, which can trigger the formation of important biomolecules on the planet \citep{Ranjan_bio_2017, rimmer_2018}. With $3^\circ$ FOV and a low-noise photon-counting detector, we will be able to detect such flares with a variability of a few milliseconds up to a few minutes \citep{Brasseur_flares_2019}. Our instrument is sensitive down to 21 AB mag, enabling us to detect even some of the faint M dwarfs. As a dedicated mission, we can also do multiple observations of the same target to obtain its light curves with high photometric accuracy.

\section{Instrument Overview}
The instrument design for {\it NUTS} was carried out keeping in mind the scientific requirements of a wide FOV with good spatial resolution while meeting the mechanical constraints of small satellites. We are nominally proposed to make use of the microsat platforms of the Indian Space Research Organization (ISRO), which gives us a constraint of fitting the science instrument in a roughly 12U volume ($300 \times 200 \times 200$ mm) with no moving parts. The tight mechanical constraints and the low UV flux levels of our observation targets limit the number of reflections in the optical path for the desired performance. A summary of the instrument parameters is given in Table.~\ref{table:instrument details}.

\begin{table*}[!ht]
\begin{center}
\begin{tabular}{ll}
\hline
\rule[-1ex]{0pt}{3.5ex} Instrument & Wide-field UV imager \\
\rule[-1ex]{0pt}{3.5ex} Telescope type & Ritchey-Chrétien  \\
\rule[-1ex]{0pt}{3.5ex} Primary aperture & 146~mm  \\
\rule[-1ex]{0pt}{3.5ex} Field of view & $3^{\circ}$ (circular)\\
\rule[-1ex]{0pt}{3.5ex} Operating bandwidth & 200~\textendash~300~nm\\
\rule[-1ex]{0pt}{3.5ex} Detector & MCP with solar-blind coating  \\
\rule[-1ex]{0pt}{3.5ex} Pore size & $10\,\mu$m \\
\rule[-1ex]{0pt}{3.5ex} Instrument Power & $ 5.5 $ W  \\
\rule[-1ex]{0pt}{3.5ex} Volume (L $\times$ W $\times$ H)& $300 \times 160\times 160$ mm  \\
\rule[-1ex]{0pt}{3.5ex} Weight & $4.5$ kg \\
\rule[-1ex]{0pt}{3.5ex} Spatial resolution & $13^{\prime\prime} $\\
\rule[-1ex]{0pt}{3.5ex} Time resolution &  2 ms full frame, 10 $\mu$s windowed \\
\rule[-1ex]{0pt}{3.5ex} Peak effective area &  12 cm$^2$ \\
\rule[-1ex]{0pt}{3.5ex} Mean effective area  &  9 cm$^2$ \\
\rule[-1ex]{0pt}{3.5ex} Sensitivity & 21 AB (with SNR of 5 in 1200 sec exposure)\\
\rule[-1ex]{0pt}{3.5ex} Bright Limit &  7 AB\\
\hline
\end{tabular}
\caption{Summary of instrument parameters}
\label{table:instrument details}
\end{center}
\end{table*}

\subsection{Optomechanical design}
\label{sec:optics}

\begin{figure}[!ht]
\begin{center}
\includegraphics[width=\linewidth]{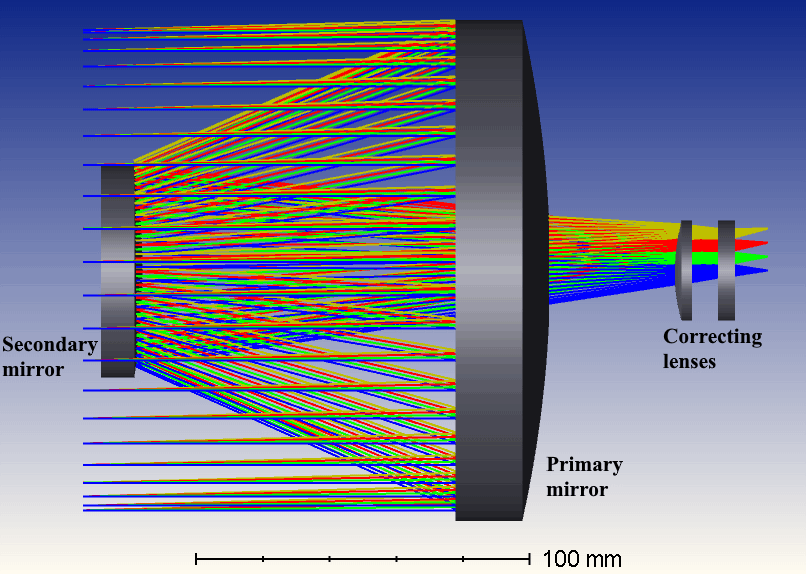} 
\end{center}
\caption{Optical layout of the instrument.}
\label{fig:layout}
\end{figure}

\begin{figure}[!ht]
\begin{center}
\frame{\includegraphics[width=\linewidth]{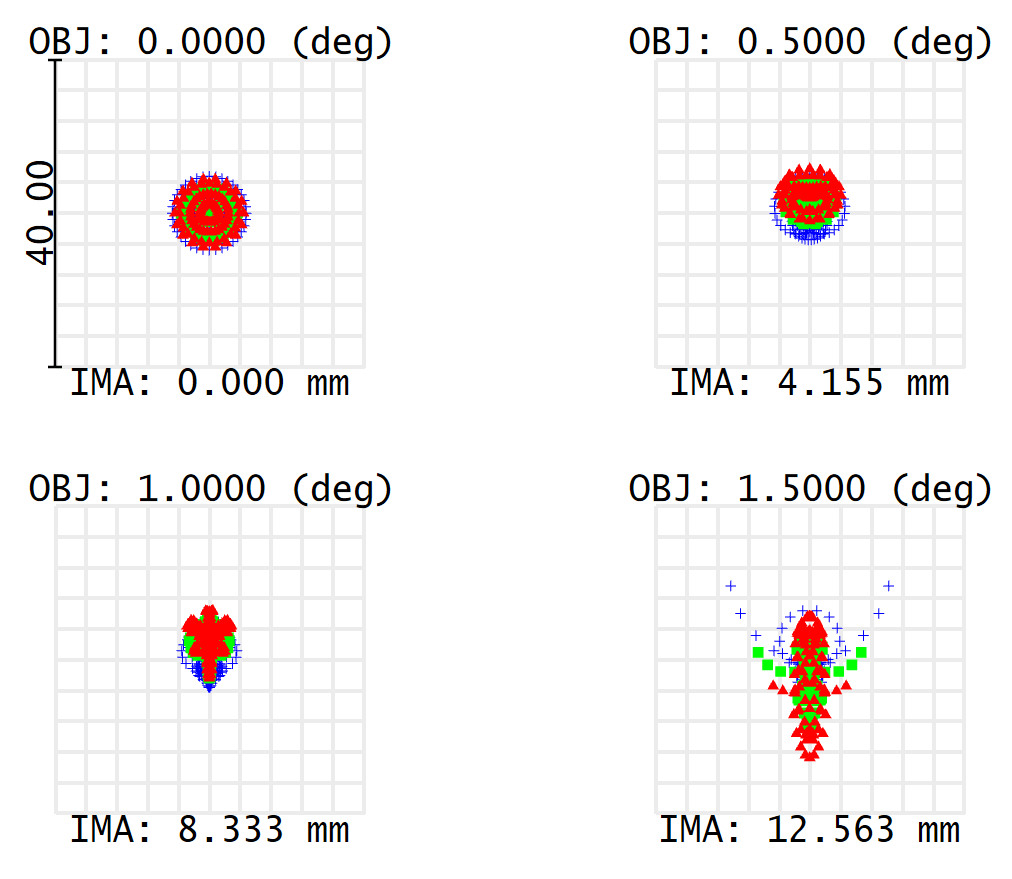}} 
\end{center}
\caption{Optical performance of the system for various field positions with the field angle (in degrees) and image position (in mm) indicated above and below the spot, respectively. The three marker colors blue, green, and red denote the instrument wavelengths of 200~nm, 250~nm, and 300~nm, respectively.}
\label{fig:spot}
\end{figure}

The optical design consists of a 150-mm primary mirror and a 64-mm secondary mirror with two 30~mm corrector lenses. Both of the mirrors are hyperbolic to eliminate the off-axis aberrations such as coma and spherical aberrations across the wide FOV, resulting in an RC  telescope design. However, the RC design and relatively large FOV introduce additional optical aberrations such as field curvature and astigmatism, which are then corrected by a set of two Fused Silica lenses. The optical layout and performance (spot diagram) are shown in Figs.~\ref{fig:layout} and \ref{fig:spot}, respectively. 

The primary and secondary mirrors were manufactured and delivered by Wavelength Opto-electronics, Singapore. They also fabricated and assembled the corrector lens assembly, with its optomechanical components. The mirrors are made of Zerodur substrate with a reflective Aluminum coating and a protective MgF\textsubscript{2} layer, giving a broad bandpass of 120~\textendash~900~nm. The set of two lenses, used to correct the aberrations, are made from fused silica with a 90\% transmission efficiency down to 195~nm. In addition, both sides of the lens are coated with an anti-reflection MgF\textsubscript{2} coating to suppress the reflection and thus maximize the transmission (up to 95\%) in the NUV band. We have also minimized the optical aberrations such that 80\% of the encircled energy falls within four pores of the MCP ($\sim$20 $\mu$m). 

\begin{figure*}[!ht]
\begin{center}
\includegraphics[width=150mm]{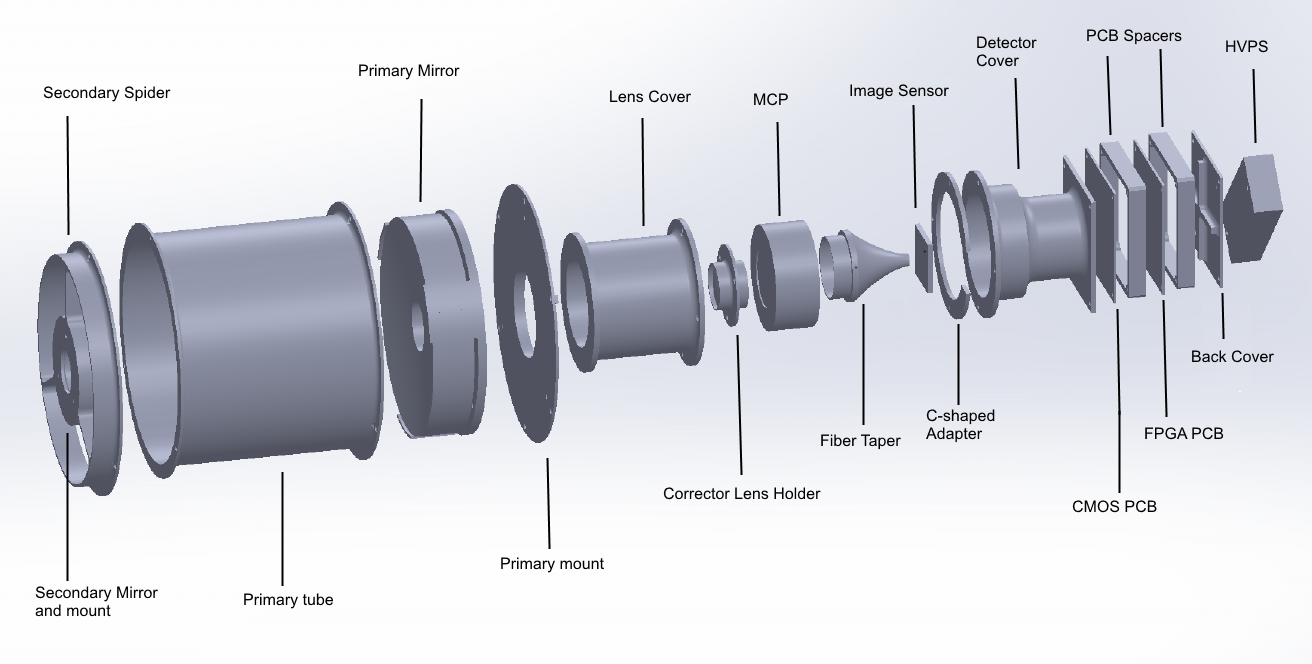} 
\end{center}
\caption{CAD model of the mechanical structure}
\label{fig:mechanical}
\end{figure*}

The mechanical design was carried out keeping in view the spacecraft requirements such as light weight, small volume, and the structural stability to withstand launch-load vibrations.  The overall design consists of the optical system, the detector unit, and the electronics box. The optical system comprises primary and secondary mirrors cells, holders for the correcting lenses, and the telescope tube that houses all these elements. Aluminum 6061-T6 is used for the tubular structure of the telescope, due to its high strength-to-density ratio, low outgassing properties, easy machinability, and low cost. The mounts for mirrors and lenses are made of Invar36 as its very low thermal expansion properties match with the Zerodur material. The optics are glued to their mounts using space grade adhesives, 3M\textsuperscript{\textregistered}~2216B. The MCP enclosure, which is attached to the lens cover, houses the MCP and the fiber optics taper. One side of the taper is glued to the phosphor side of the MCP and the other to the CMOS. The electronic box consists of the a series of three PCBs and the high voltage supply box. An exploded CAD drawing of the instrument is shown in Fig.~\ref{fig:mechanical}.

The instrument is attached to the spacecraft with an interface flange at the center of gravity (CG) for better mechanical stability. From our vibration analyses and launch provider specificatios \citep{Mathew_2017, Kumar_UVIT_2012}, we have determined that the natural frequency must be above 100~Hz, and the instrument should be able to withstand shock loads of up to 5 g. We have also taken into account the manufacturing and alignment tolerances and temperature fluctuations that affect the final Point Spread Function (PSF) to ensure the performance for even the widest FOV. A baseline performance estimate ensuring that 80\% encircled energy falls within a $20\,\mu$m radius was adopted, and the tolerance analysis was done using Zemax Optics Studio. The tolerance limits were derived (Table~\ref{table:tolerance}) using the telescope focus as the compensator, which can be adjusted with the help of shims during the calibration. From here, we also concluded that an operating temperature of $15^\circ$~\textendash~$25^\circ$C is optimal for the optomechanical unit to give the desired performance, which can be achieved by both passive (multi-layer insulation) and active (heaters) methods.  

\begin{table}[!ht]
\begin{center}
\small
\begin{tabular}{lll}
\hline 
\rule[-1ex]{0pt}{3.5ex} Tolerance type & Tolerance term & Value \\
\hline
\rule[-1ex]{0pt}{3.5ex} Manufacture & Radius of Curvature (Mirror) & 1~\% \\
\rule[-1ex]{0pt}{3.5ex} & Radius of Curvature (Lens) & 0.1~\% \\
\rule[-1ex]{0pt}{3.5ex} & Thickness & 50$~\mu$m \\
\rule[-1ex]{0pt}{3.5ex} & X-Y decenter & 50$~\mu$m \\
\rule[-1ex]{0pt}{3.5ex} & X-Y tilt & $60^{\prime\prime}$ \\
\rule[-1ex]{0pt}{3.5ex} & Surface roughness & $\lambda$/6\\
\rule[-1ex]{0pt}{3.5ex} Alignment & X-Y decenter & 50$~\mu$m \\
\rule[-1ex]{0pt}{3.5ex} & X-Y tilt & $60^{\prime\prime}$ \\
\hline
\end{tabular}
\label{table:tolerance}
\end{center}
\caption{Manufacture and alignment tolerances for the instrument.}
\end{table}

\begin{figure}[!ht]
\begin{center}
\includegraphics[width=\linewidth]{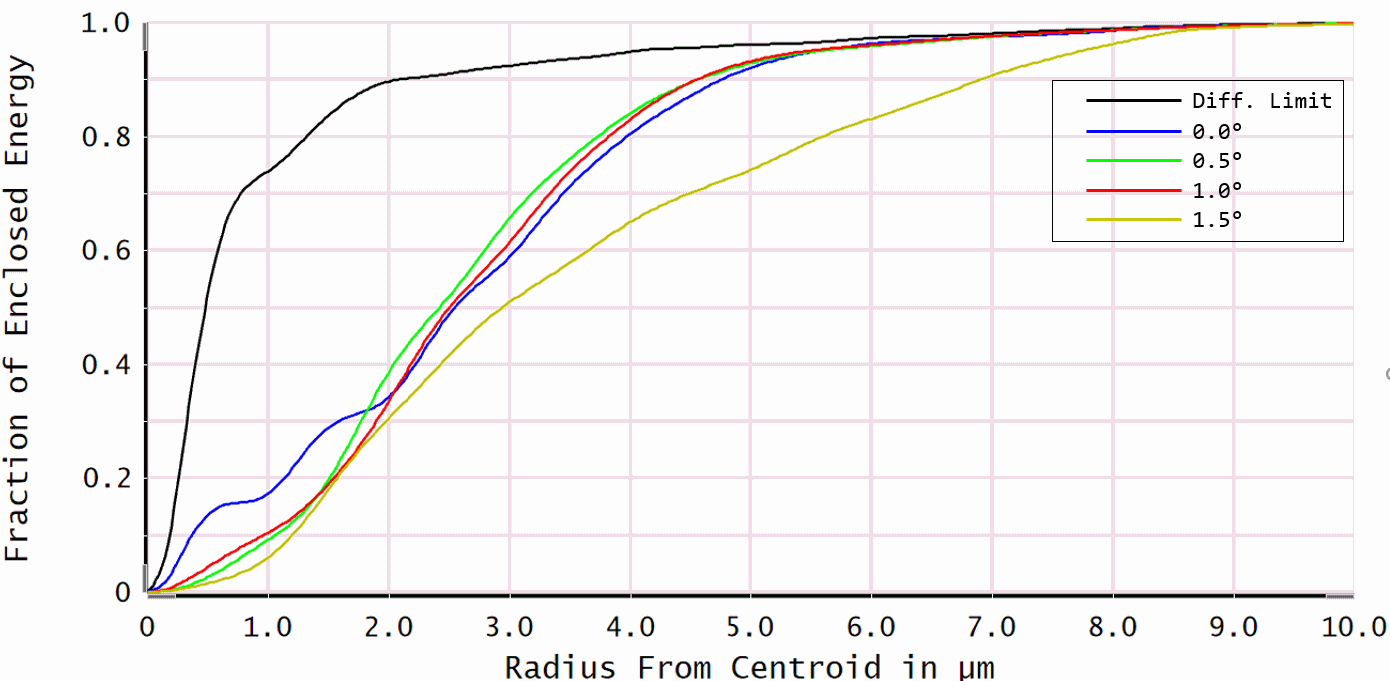} 
\end{center}
\caption{Diffraction-encircled energy for on-axis and off-axis fields. The colours indicate different fields, with the percentage of energy distribution (in $y$ axis) as a function of the distance from the spot center (in $x$ axis).}
\label{fig:enc}
\end{figure}

\subsection{Detector and electronics}
\label{sec:detector}

\begin{figure*}[!ht]
\begin{center}
\includegraphics[width=\linewidth]{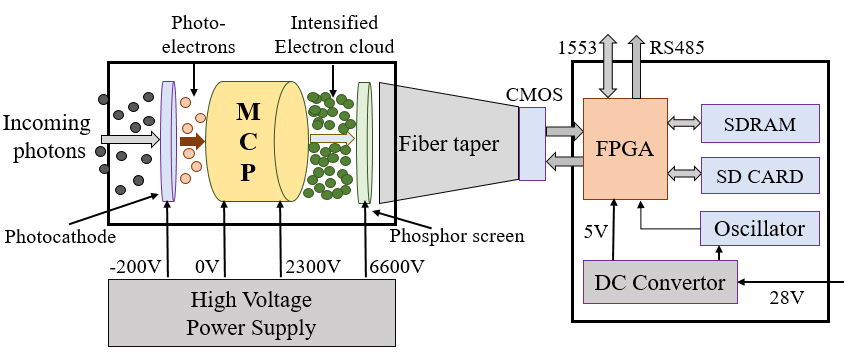} 
\end{center}
\caption{Block diagram of the detector system. The electrons from the MCP are converted to green photons by the phosphor screen at its output. This light is coupled to a CMOS chip using a fiber taper. }
\label{fig:detector_diagram}
\end{figure*}

Photon-counting detectors using MCPs have been the standard for UV payloads because of their low readout noise, large detector area, radiation tolerance, and long-wavelength rejection \citep{Kimble_2003_SPIE, Joseph_1995}. We have achieved significant savings in cost by using a phosphor-screen anode as the readout, and by focusing the image of the anode onto a CMOS sensor as shown in Fig.~\ref{fig:detector_diagram}. We have developed and tested a laboratory model of the detector \citep{Ambily_2017} for some of our balloon payloads \citep{NUV_Spectrograph_Sreejith, Mathew_2018}.  The detector for {\it NUTS} is a modified version of this with a solar blind photocathode made of Cesium Telluride to address the red leak and dark current issues of the S20 photocathode. As most of the UV targets are relatively faint, the detector system works in the photon-counting mode, where it records the position and arrival time of each photon with an accuracy of $13^{\prime\prime}$ and 2 ms, respectively. The key part of the detector is the MCP340 assembly from Photek\footnote{\tt{http://www.photek.com/}}. It is a 40-mm
Z-stack MCP with a solarblind photocathode deposited on a fused Silica input window and a P46 phosphor deposited on the anode. The gain of the MCP is varied by the voltage at the MCP output, which automatically adjusts the screen voltage, maintaining a constant voltage difference between the Anode and MCP\_Out terminals. The resulting electron cloud from the MCP output electrode is accelerated towards the phosphor screen anode (with a peak response at 530~nm), and the resulting photons are focused on the CMOS surface by a custom-made fiber taper, also manufactured by Photek. 

Since we are interested in time-domain astronomy, it is important to have a CMOS image sensor and readout electronics that can support high frame rates. The LUPA1300-2 (NOIL2SM1300A) is a 1.6 MP sensor from ON  Semiconductor that meets our requirements of faster readout and a large sensor size. It has 14 $\mu$m pixels with $18 \times 14$ mm size, which matches the output diameter of the taper, and gives rise to fewer aberrations when interfaced with the MCP. It also supports high-speed clocks of up to 315 MHz rendering a maximum frame transfer rate of up to 500 fps. The sensor further supports windowed and sub-sampled readouts, which means we can achieve a temporal resolution of 2 ms for the full resolution of 1.3 MP and 10 $\mu$s for the smallest window. We have designed a custom readout board to give the necessary bias and clock lines for the sensor. It also interfaces the synchronization clocks and pixel values to the FPGA board. The data acquisition and processing unit is a field-programmable gate array (FPGA) board, which is a generic board we have developed to be used with multiple CMOS and CCD sensors \citep{Sarpotdar_2016_SPIE}. It is based on a MIL-grade Spartan-6Q FPGA from Xilinx\footnote{Xilinx Inc.: \tt{http://www.xilinx.com/}} with a wide operating temperature range. This digital readout board also performs the necessary image processing operations for the photon-counting mode, including computing the centroids of photon events to sub pixel accuracy \citep{Ambily_2017}. The board contains the necessary peripherals for storage of the intermediate data (SDRAM, Flash, and SD Card) and interfacing with other spacecraft electronics (UART, I\textsuperscript{2}C, and JTAG). 

\subsubsection{Photon event centroiding}

The electron cloud from the MCP is roughly Gaussian in shape \citep{Joseph_1995, Hutchings_2007}. To accurately determine the $X-Y$ position of the photon on the photocathode, we need to determine the centroid of the resulting photon cloud on the CMOS to sub-pixel accuracy. The centroiding algorithm was implemented using a $5\times 5$-pixel sliding window, to first detect a possible photon even which would appear as a local maximum in the window. We estimate the background level in each window from the lowest corner pixel value \citep{{Hutchings_2007},{Ambily_2017}}, which is then used to flag hot pixels and multiple events. Once the actual photon event locations are identified, the subpixel or fractional position values are computed by a weighted mean algorithm. In the centroiding mode, the output data for telemetry is in the form of packets. Each packet contains a frame ID, event ID, integer and fractional values of centroid coordinates, and the intensity value of the central pixel. 

\begin{figure*}[!ht]
\begin{center}
\includegraphics[width=150mm]{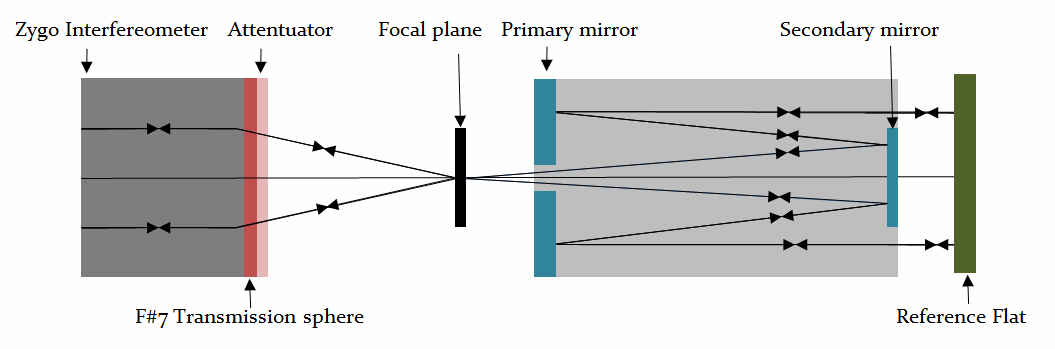} 
\end{center}
\caption{Fine alignment setup using Zygo interferometer. The mechanical axis of the system was first established with a theodolite, which was then used by the interferometric setup to align the primary and secondary mirrors with the optical axis of the telescope. }
\label{fig:nuts_alignment}
\end{figure*}
\section{Assembly and Calibration}
\label{sec:cal}

The assembly, alignment and initial calibration of the instrument were performed at the M.~G.~K.~Menon Laboratory (MGKML) for Space Sciences at IIA. We have employed the procedures and techniques described in \cite{Mathew_2019} for the assembly and calibration of the Lunar Ultraviolet Cosmic Imager ({\it LUCI}, \citet{Mathew_2017}). This facility was also used by the {\it UVIT} payload on {\it AstroSat} \citep{Kumar_UVIT_2012} and includes the necessary equipment for calibrating NUV instruments while ensuring cleanliness. We have access to class 10000 cleanrooms with laminar flow tables providing a class 1000 or better local environment. 


\subsection{Optical alignment and calibration}

We have performed the coarse alignment of the telescope using a theodolite and a pair of cross-hair targets on a vibration controlled optical table. The theodolite has a precision of up to 0.01\as, which was used to establish the mechanical axis of the system. This was used to align the optical axis of the system with the help of a Zygo interferometer and a reference flat mirror. As shown in Fig.~\ref{fig:nuts_alignment}, this setup was used to align the primary and secondary optics of the system to the opto-mechanical axis.

The calibration steps include the measurement of system throughput, and the optical performance estimates, using the same equipment and methods described in \cite{Mathew_2019}. UV light from a deuterium lamp is coupled to a collimator setup through an Acton monochromator, which will be used along with NIST-calibrated photodiodes for the effective area measurements. The optical performance will be estimated by imaging a set of pinhole masks of varying sizes from different points in the field.

Once the full ground calibration and qualification of the payload are completed, we plan to commence the environmental tests at the MGKML. The instrument will undergo testing for vibration performance, thermal vacuum tests including thermal balance and thermal cycling, and EMI/EMC tests. Being a UV telescope, {\it NUTS} is extremely sensitive to contamination, and all the components are cleaned using bake-out, and ultrasonic and solvent cleansing. The instrument and individual flight components are also kept in a nitrogen purge during and after the tests. 

\section{Performance estimates}

\begin{figure}[!ht]
\begin{center}
\includegraphics[width=\linewidth]{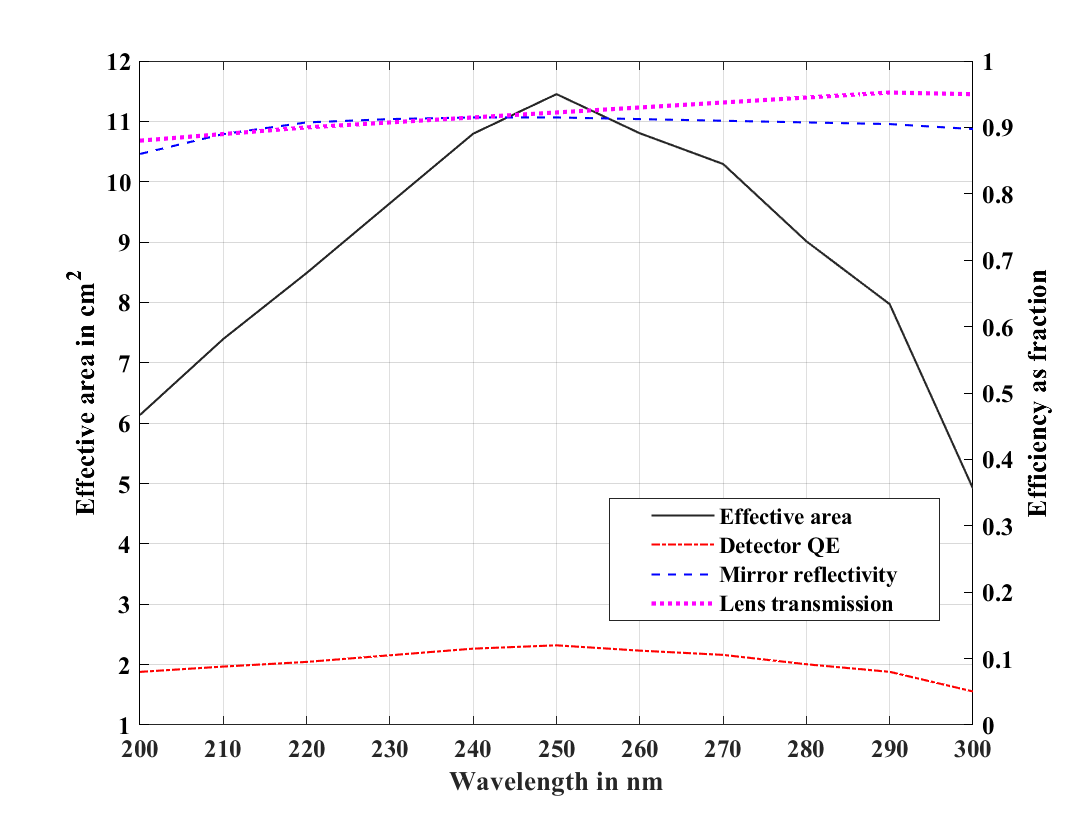} 
\end{center}
\caption{Estimated effective area of the instrument (black solid line) on the left axis. On the right axis, the mirror reflectivity values are measured and provided by the manufacturer (blue dashes), lens transmission was measured in the lab (purple dots), and the detector quantum efficiency (red dot-dashes) was estimated from the manufacturer's numbers for the solar blind photocathode and from our lab measurements.}
\label{fig:nuts_ea}
\end{figure}

We have estimated the system throughput and instrument effective area from the components level tests, by using the mirror reflectivity values, and the measured values for lens transmission and detector quantum efficiency. The total system response in terms of effective area in cm$^2$ can be expressed as
\begin{equation}
A_{\rm eff}
= A_{\rm coll}\times R_{\rm PM}(\lambda) \times R_{\rm SM}(\lambda) \times T_{\rm l}^{2} (\lambda) \times {\rm QE}(\lambda)\,,
\end{equation}
where $A_{\rm coll}$ is the effective geometrical collecting area, $R_{\rm PM}$ and $R_{\rm SM}$ are the reflectivity of the primary and the secondary mirrors respectively, $T_{\rm l}$ is the lens transmission, and QE$(\lambda)$ is the quantum efficiency of the detector. The effective area plot for the instrument is shown in Fig.~\ref{fig:nuts_ea}, which shows that the instrument has a peak effective area of 12 cm$^2$.

\begin{figure*}[!ht]
\begin{center}
\includegraphics[width=120mm]{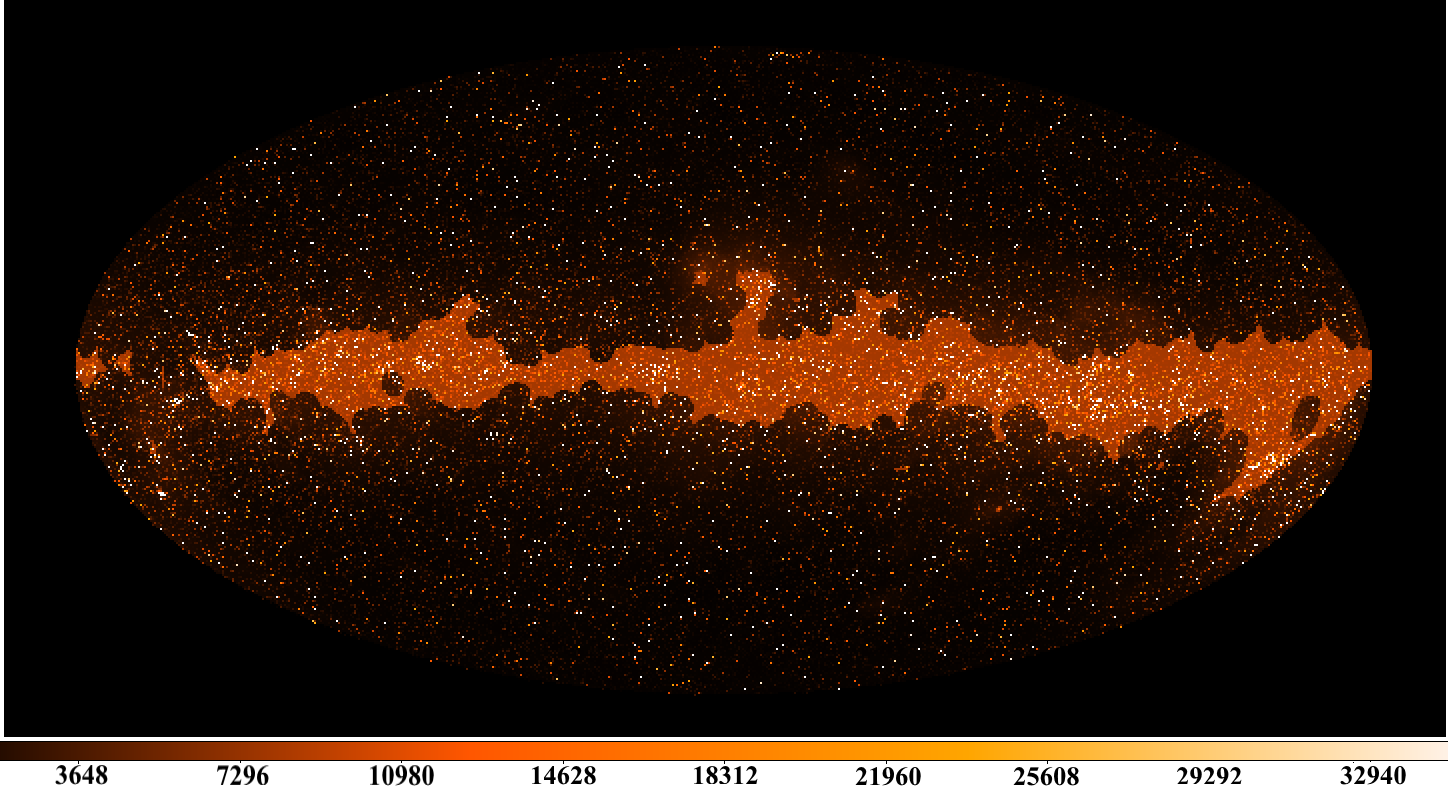} 
\end{center}
\caption{Simulated UV Sky as seen by {\it NUTS}. The color bar represents the estimated counts, which include the diffuse background, zodiacal light, and stellar contributions taken from archival observations. }
\label{fig:nuts_sky}
\end{figure*}

We have also used the UV sky simulator \citep{safonova_simulator_2013} to model the UV sky as could be seen by the instrument. The tool takes into account diffuse and point sources as well as the instrumental background estimates. The point sources accounted for here are the bright stars from the Hipparcos catalog, whose flux in $200~\textendash~300$~nm range and the count rates are estimated from their model spectra. The diffuse components include the NUV background, zodiacal light (for January in this case), and airglow contribution. These diffuse observations are taken from {\it GALEX}, which means that the galactic plane remains unobserved. The diffuse background in these regions is estimated by applying cosecant law \citep{Murthy_cosecant_2010}. The estimated flux values are converted to instrument counts using the effective area curve from Fig.~\ref{fig:nuts_ea}, assuming zero dark current. These numbers are combined to generate an output file which gives the total count rate in a given $3^\circ$ field. The all-sky map as could be seen by {\it NUTS} is given in Fig.~\ref{fig:nuts_sky}.

\begin{figure*}[!ht]
\begin{center}
\includegraphics[width=\textwidth]{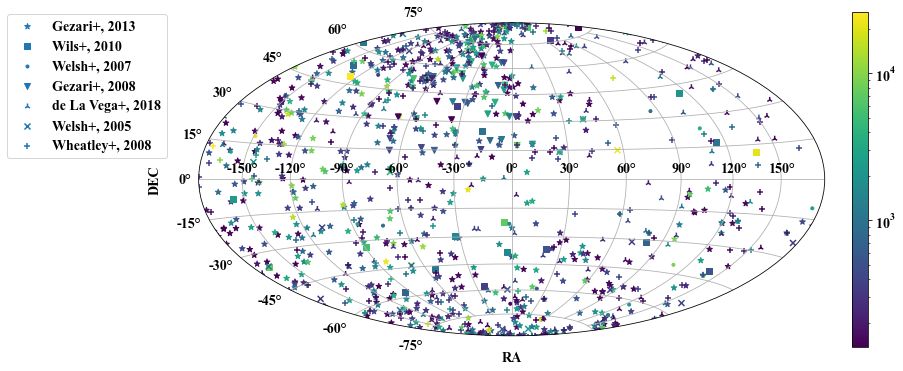} 
\end{center}
\caption{Variable sources in {\it GALEX} catalogs as seen by {\it NUTS}, scattered along their RA and DEC. The colormap represents the estimated counts. The NUV magnitudes from these variability catalogs were converted to flux using {\it GALEX} conversion factors. These flux values, along with the estimated throughput of {\it NUTS}, were used to determine the photon counts as seen by the instrument in 1200 sec.}
\label{fig:nuts_variable_sky}
\end{figure*}

To determine the feasibility of using {\it NUTS} for temporal variability studies, we also looked at previous observations of transient events in archival data, particularly from the {\it GALEX} mission. We made use of the observations from the variability catalogs \citep{Welsh_variability_2005, wheatley_varability_2008} and time domain survey \citep{galex_gezari_2013}, as well as additional catalogs of novae outbursts \citep{Wils_novae_2010} and M dwarf flares \citep{Welsh_mdwarf_2007}. In addition, there have been new archival studies of short timescale UV variability \citep{delavega_variability_2018, Brasseur_flares_2019} using the gPhoton package \citep{million_gphoton_2017}. These catalogs were used to determine the median NUV flux levels of these events, which were then converted to their equivalent counts as would be seen by NUTS. The image in Fig.~\ref{fig:nuts_variable_sky} shows one subset of such observations, where the estimated counts for {\it NUTS} are shown across the sky and shows that we would have the possibility of detecting UV transients across various pointing directions and magnitude ranges.

\section{Launch and Operations plan}

We are proposing to fly \textit{\textbf{NUTS}} on the small satellite platform (PS4-OP) of the Indian Space Research Organization (ISRO) to a polar low-Earth orbit \citep{isro_ao}. The platform provides the electrical and data interfaces via an IEEE 1553 interface, in addition to a 28~V  power supply for the electronics box. The Attitude Determination and Control System (ADCS) system includes a star tracker capable of providing a pointing accuracy of $30^{\prime\prime}$ and stability of about $1^{\prime\prime}$ per second. The mechanical interface is provided as a flange at the center of gravity of the instrument. The spacecraft is also estimated to actively maintain the instrument temperature within $\pm10^\circ$C, with the help of temperature sensors and heaters attached to the instrument body. 

Our default observation mode as a transient survey telescope, is to stare at the UV sky in an anti-solar direction. The science observations happen during the orbital night while maintaining the telescope's Moon and Earth avoidance angles at $40^{\circ}$ and $25^{\circ}$ respectively. Each pointing will cover a 7 square degrees field and last 1200 seconds, which gives us a limiting AB magnitude of 21. Further calculations are done to ensure that we avoid brighter targets ($< 7$th AB) to preserve detector life. 
The expected data rate for the instrument is 100 kbps, with data consisting of the photon event lists logged during the observations. The events are registered as individual photons on the detector which will be assembled into a data stream containing the $X-Y$ position and pulse height and transmitted to the ground, along with time tags. As we are going to be in LEO  and observations occur only in eclipse, we have about 25 minutes of observing time per orbit. This amounts to a maximum science data volume of $\sim$32 MB per day, with a 5-byte data packet per photon event. We plan to implement further data compression algorithms to reduce the data volume to around 5 to 10~MB/day. \textit{\textbf{NUTS}} is also capable of doing longer observations of any potentially interesting transients, by pointing the telescope for multiple orbits.

One of the crucial tasks for the mission now is the development of the data pipeline and in-flight calibration techniques. After receiving the photon lists at the ground station, an automated pipeline will convert it into a 2D image. The pipeline software will draw on the previous software developed at IIA for missions such as {\it TAUVEX} \citep{rekhesh_tauvex_2008} and {\it UVIT} \citep{Murthy_jude_2017}. The photometric calibration of the instrument will also follow the procedures established by previous UV space missions such as {\it HST} \citep{bohlin_hst_1986}, {\it GALEX} \citep{galex_morrisey_2007}, and {\it UVIT} \citep{UVIT_rahna_2017}. We will use a set of well studied standard stars, such as the UV-bright white dwarfs LB227 and HZ4, for the in-flight calibration of the instrument \citep{bohlin_calibrations_2014}. The ground and on-orbit calibrations will help us identify the photometric errors as a function of NUV magnitude, which we would then use to set the criteria for a transient event (say, $>5 \sigma$). These open source software would be made available to the community, to generate calibrated light curves from the photon event lists.

\section{Summary and Conclusions}
We have described the science motivation, instrument design, and calibration procedures for the wide-field NUV imager, {\it NUTS}. The mission attempts to fill the gap in available UV resources and would demonstrate the capabilities of a wide field camera for UV transient surveys. Currently, we have designed, fabricated, and assembled the components, and the final calibrations and tests are being carried out currently. Once in orbit, the instrument will stare at large areas of the sky and the resulting data will be monitored in real-time to initiate follow-up observations for interesting transients. This will also be followed up by ground-based optical telescopes, such as e.g. the Himalayan Chandra Telescope (HCT) of the Indian Astronomical Observatory (IAO), Leh, Ladakh. 

\section{Acknowledgments}

Part of this research has been supported by the Department of Science and Technology (Government of India) under Grants IR/S2/PU-006/2012 and EMR/2016/001450/PHY. This project was funded in part by the Austrian Science Fund (FWF) [J 4596-N]. MS acknowledges the financial support by the Department of Science and Technology (DST), Government of India, under the Women Scientist Scheme (PH), project reference number SR/WOS-A/PM-17/2019. We thank Mr. S. Sriram and all the staff at the M.~G.~K.~Menon laboratory (CREST) for helping us with optics assembly and storage of payload components in the cleanroom environment.



\begin{thebibliography}{99}
\bibitem[Ambily et al.(2017)]{Ambily_2017} Ambily, S., Sarpotdar, M., Mathew, J., et al.\ 2017, ``Development of Data Acquisition Methods for an FPGA-Based Photon Counting Detector", \jai, 6, 1750002 
\bibitem[Bellm et al.(2019)]{ZTF_bellm_2019} Bellm, E.~C., Kulkarni, S.~R., Graham, M.~J., et al.\ 2019, \pasp, 131, 018002. doi:10.1088/1538-3873/aaecbe
\bibitem[Bianchi et al.(2017)]{bianchi_GalCat_2017} Bianchi, L., Shiao, B., \& Thilker, D.\ 2017, \apjs, 230, 24. doi:10.3847/1538-4365/aa7053
\bibitem[Bohlin et al.(2014)]{bohlin_calibrations_2014} Bohlin, R.~C., Gordon, K.~D., \& Tremblay, P.-E.\ 2014, \pasp, 126, 711. doi:10.1086/677655
\bibitem[Brasseur et al.(2019)]{Brasseur_flares_2019} Brasseur, C.~E., Osten, R.~A., \& Fleming, S.~W.\ 2019, \apj, 883, 88. doi:10.3847/1538-4357/ab3df8
\bibitem[Brosch, Balabanov \& Behar(2014)]{Noah_Cubesat_2014} Brosch, N., Balabanov, V. \& Behar, E. 2014, ``Small observatories for the UV", \apss, 355, 205
\bibitem[Cenko(2019)]{Cenko_Dorado_2019} Cenko, S.~B.\ 2019, \aas
\bibitem[Chandra et al.(2020)]{bharat_sing_2020} Chandra, B., Sachkov, M., Prabha, S., et al.\ 2020, \procspie, 11444, 1144476. doi:10.1117/12.2563051
\bibitem[de la Vega \& Bianchi(2018)]{delavega_variability_2018} de la Vega, A. \& Bianchi, L.\ 2018, \apjs, 238, 25. doi:10.3847/1538-4365/aaddf5
\bibitem[Dressing \& Charbonneau(2015)]{dressing_2015} Dressing, C.~D. \& Charbonneau, D.\ 2015, \apj, 807, 45. doi:10.1088/0004-637X/807/1/45
\bibitem[Fleming et al.(2018)]{cute_fleming_2018} Fleming, B.~T., France, K., Nell, N., et al.\ 2018, ``Colorado Ultraviolet Transit Experiment: a dedicated CubeSat mission to study exoplanetary mass loss and magnetic fields", \jatis, 4, 014004 
\bibitem[Fleming et al.(2019)]{Fleming_sprite_2019} Fleming, B.~T., France, K., Williams, J., et al.\ 2019, \procspie, 11118, 111180U. doi:10.1117/12.2529512
\bibitem[France et al.(2016)]{france_rockets_2016} France, K., Hoadley, K., Fleming, B.~T., et al.\ 2016, ''The SLICE, CHESS, and SISTINE Ultraviolet Spectrographs: Rocket-Borne Instrumentation Supporting Future Astrophysics Missions", \jai, 5, 1640001 
\bibitem[Ganot et al.(2016)]{ganot_SN_2016} Ganot, N., Gal-Yam, A., Ofek, E.~O., et al. "The Detection Rate of Early UV Emission from Supernovae: A Dedicated Galex/PTF Survey and Calibrated Theoretical Estimates'', \ 2016, \apj, 820, 57 
\bibitem[Gaudi et al.(2020)]{habex_report_2020} Gaudi, B.~S., Seager, S., Mennesson, B., et al.\ 2020, arXiv:2001.06683
\bibitem[Gezari et al.(2013)]{galex_gezari_2013} Gezari, S., Martin, D.~C., Forster, K., et al.\ 2013, \apj, 766, 60. doi:10.1088/0004-637X/766/1/60
\bibitem[Green et al.(2012)]{cos_green_2012} Green, J.~C., Froning, C.~S., Osterman, S., et al.\ 2012, ''The Cosmic Origins Spectrograph", \apj, 744, 60 
\bibitem[Howard et al.(2012)]{Howard_exoplanets_2012} Howard, A.~W., Marcy, G.~W., Bryson, S.~T., et al.\ 2012, \apjs, 201, 15. doi:10.1088/0067-0049/201/2/15
\bibitem[Hutchings et al.(2007)]{Hutchings_2007} Hutchings, J.~B., Postma, J., Asquin, D., et al.\ 2007, \pasp, 119, 1152. doi:10.1086/522635
\bibitem[PS4-OP(2019)]{isro_ao} doi:http://www.isro.gov.in/sites/default/files/orbital\_
platform-\_ao.pdf
\bibitem[Ivezi{\'c} et al.(2019)]{lsst_2019} Ivezi{\'c}, {\v{Z}}., Kahn, S.~M., Tyson, J.~A., et al.\ 2019, \apj, 873, 111. doi:10.3847/1538-4357/ab042c
\bibitem[Johnstone et al.(2015)]{johnstone_hab_2015} Johnstone, C.~P., G{\"u}del, M., St{\"o}kl, A., et al.\ 2015, \apjl, 815, L12. doi:10.1088/2041-8205/815/1/L12
\bibitem[Joseph(1995)]{Joseph_1995} Joseph, C.~L.\ 1995, ''UV Image Sensors and Associated Technologies", \ea, 6, 97 
\bibitem[Kimble et al.(2003)]{Kimble_2003_SPIE} Kimble, R.~A., Pain, B., Ortiz, M., et al.\ 2003, ''A high-speed event-driven active pixel sensor readout for photon-counting microchannel plate detectors", \procspie, 4854, 203 
\bibitem[Kumar et al.(2012)]{Kumar_UVIT_2012} {Kumar}, A., {Ghosh}, S.~K., {Hutchings}, J., et al. 2012, ''Ultra Violet Imaging Telescope (UVIT) on ASTROSAT", \procspie, 8443, 84431N 
\bibitem[Law et al.(2009)]{PTF_law_2009} Law, N.~M., Kulkarni, S.~R., Dekany, R.~G., et al.\ 2009, \pasp, 121, 1395. doi:10.1086/648598
\bibitem[Martin et al.(2005)]{Galex_martin_2005} Martin, D.~C., Fanson, J., Schiminovich, D., et al.\ 2005, ''The Galaxy Evolution Explorer: A Space Ultraviolet Survey Mission", \apj, 619, L1
\bibitem[Mason et al.(2001)]{Mason_xmm_2001} Mason, K.~O., Breeveld, A., Much, R., et al.\ 2001, ''The XMM-Newton optical/UV monitor telescope", \aap, 365, L36 
\bibitem[Mathew et al.(2017)]{Mathew_2017} Mathew, J., Prakash, A., Sarpotdar, M., et al.\ 2017, ''Prospect for UV observations from the Moon. II. Instrumental Design of an Ultraviolet Imager LUCI", \apss, 362, 37 
\bibitem[Mathew et al.(2018a)]{Mathew_2018} Mathew, J., Ambily, S., Prakash, A., et al.\ 2018, ''Wide-field Ultraviolet Imager for Astronomical Transient Studies", \ea, 45, 201 
\bibitem[Mathew et al.(2019)]{Mathew_2019} Mathew, J., Nair, B.~G., Safonova, M., et al.\ 2019, \apss, 364, 53. doi:10.1007/s10509-019-3538-8
\bibitem[Million et al.(2016)]{million_gphoton_2017} Million, C., Fleming, S.~W., Shiao, B., et al.\ 2016, \apj, 833, 292. doi:10.3847/1538-4357/833/2/292
\bibitem[Mohan(2008)]{rekhesh_tauvex_2008} Mohan, R.\ 2008, ''TAUVEX observatory activities - Software tools", {\it Bulletin of the Astronomical Society of India Proceedings}, 25, 77 
\bibitem[Moos et al.(2000)]{fuse_moos_2000} Moos, H.~W., Cash, W.~C., Cowie, L.~L., et al.\ 2000, ''Overview of the Far Ultraviolet Spectroscopic Explorer Mission", \apjl, 538, L1 
\bibitem[Morrissey et al.(2007)]{galex_morrisey_2007} Morrissey, P., Conrow, T., Barlow, T. et al., 2007,  ``The on-orbit performance of the galaxy evolution explorer", \apjs, 173, 682
\bibitem[Murthy et al.(2010)]{Murthy_cosecant_2010} Murthy, J., Henry, R.~C., \& Sujatha, N.~V.\ 2010, \apj, 724, 1389. doi:10.1088/0004-637X/724/2/1389
\bibitem[Murthy et al.(2017)]{Murthy_jude_2017} Murthy, J., Rahna, P.~T., Sutaria, F., et al.\ 2017, ''JUDE: An Ultraviolet Imaging Telescope pipeline", {\it Astronomy and Computing}, 20, 120 \bibitem[Bohlin(1986)]{bohlin_hst_1986} Bohlin, R.~C.\ 1986, ''The ultraviolet calibration of the Hubble Space Telescope. I - Secondary standards of absolute ultraviolet flux and the recalibration of IUE", \apj, 308, 1001 
\bibitem[Nakar \& Sari(2010)]{SN_UV_NakarSari_2010} Nakar, E. \& Sari, R.\ 2010, \apj, 725, 904. doi:10.1088/0004-637X/725/1/904
\bibitem[Oke \& Gunn(1983)]{OkeGunn_AB} Oke, J.~B., \& Gunn, J.~E.\ 1983,  {\apj}, 266, 713
\bibitem[Rabinak \& Waxman(2011)]{Rabinak_2011} Rabinak, I. \& Waxman, E.\ 2011, \apj, 728, 63. doi:10.1088/0004-637X/728/1/63
\bibitem[Rahna et al.(2017)]{UVIT_rahna_2017} Rahna, P.~T., Murthy, J., Safonova, M., et al.\ 2017, ''Investigating the in-flight performance of the UVIT payload on AstroSat", \mnras, 471, 3028 
\bibitem[Ranjan et al.(2017)]{Ranjan_bio_2017} Ranjan, S., Wordsworth, R., \& Sasselov, D.~D.\ 2017, \apj, 843, 110. doi:10.3847/1538-4357/aa773e
\bibitem[Ridden-Harper et al.(2017)]{gluv_harper_2017} Ridden-Harper, R., Tucker, B.~E., Sharp, R., Gilbert, J., \& Petkovic, M.\ 2017, ''Capability of detecting ultraviolet counterparts of gravitational waves with GLUV", \mnras, 472, 4521 
\bibitem[Rimmer et al.(2018)]{rimmer_2018} Rimmer, P.~B., Xu, J., Thompson, S.~J., et al.\ 2018, Science Advances, 4, eaar3302. doi:10.1126/sciadv.aar3302
\bibitem[Roming et al.(2005)]{uvot_roming_2005} Roming, P.~W.~A., Kennedy, T.~E., Mason, K.~O., et al.\ 2005, ''The Swift Ultra-Violet/Optical Telescope", \ssr, 120, 95 
\bibitem[Rugheimer et al.(2015)]{Rugheimer_UVrad_2015} Rugheimer, S., Kaltenegger, L., Segura, A., et al.\ 2015, \apj, 809, 57. doi:10.1088/0004-637X/809/1/57
\bibitem[Safonova et al.(2013)]{safonova_simulator_2013} Safonova, M., Mohan, R., Sreejith, A.~G., \& Murthy, J.\ 2013, ''Predicting UV sky for future UV missions", {\it Astronomy and Computing}, 1, 46 
\bibitem[Safonova et al.(2014)]{safonova_luci_2014} Safonova, M., Mathew, J., Mohan, R., et al.\ 2014, ''Prospect for UV observations from the Moon", \apss, 353, 329 
\bibitem[Sagiv et al.(2014)]{ultrasat_sagiv_2014} Sagiv, I., Gal-Yam, A., Ofek, E.~O., et al.\ 2014 , ``Science with a Wide-field UV Transient Explorer", \aj, 147, 79
\bibitem[Sarpotdar et al.(2016)]{Sarpotdar_2016_SPIE} Sarpotdar, M., Mathew, J., Safonova, M., \& Murthy, J.\ 2016, ''A generic FPGA-based detector readout and real-time image processing board", \procspie, 9915, 99152K 
\bibitem[Shkolnik et al.(2018)]{sparcs_shkolnik_2018} Shkolnik, E.~L., Ardila, D., Barman, T., et al.\ 2018, ''Monitoring the High-Energy Radiation Environment of Exoplanets Around Low-mass Stars with SPARCS (Star-Planet Activity Research CubeSat)", American Astronomical Society Meeting Abstracts, 231, 228.04 
\bibitem[Sreejith et al.(2015)]{NUV_Spectrograph_Sreejith} Sreejith, A.~G., Safonova, M., \& Murthy, J.\ 2015, ''Near ultraviolet spectrograph for balloon platform", \procspie, 9654, 96540D \bibitem[Sreejith et al.(2016)]{Sreejith_balloon_2016} Sreejith, A.~G., Mathew, J., Sarpotdar, M., et al.\ 2016, ''Balloon UV Experiments for Astronomical and Atmospheric Observations", \procspie 9908, 99084E
\bibitem[Tandon et al.(2017)]{Tandon_UVIT_2017} Tandon, S.~N., Subramaniam, A., Girish, V., et al.\ 2017, \aj, 154, 128. doi:10.3847/1538-3881/aa8451
\bibitem[The LUVOIR Team(2019)]{luvoir_report_2019} The LUVOIR Team\ 2019, arXiv:1912.06219
\bibitem[Wang et al.(2019)]{wang_2019} Wang, J., Liang, E.~W., \& Wei, J.~Y.\ 2019, \pasp, 131, 095001. doi:10.1088/1538-3873/ab2749
\bibitem[Wang et al.(2019)]{Wang_cl-agn_2019} Wang, J., Xu, D.~W., Wang, Y., et al.\ 2019, \apj, 887, 15. doi:10.3847/1538-4357/ab4d90
\bibitem[Welsh et al.(2005)]{Welsh_variability_2005} Welsh, B.~Y., Wheatley, J.~M., Heafield, K., et al.\ 2005, \aj, 130, 825. doi:10.1086/431222
\bibitem[Welsh et al.(2007)]{Welsh_mdwarf_2007} Welsh, B.~Y., Wheatley, J.~M., Seibert, M., et al.\ 2007, \apjs, 173, 673. doi:10.1086/516640
\bibitem[Welsh et al.(2011)]{welsh_SN_2011} Welsh, B.~Y., Wheatley, J.~M., \& Neil, J.~D.\ 2011, \aap, 527, A15. doi:10.1051/0004-6361/201015865
\bibitem[Wheatley et al.(2008)]{wheatley_varability_2008} Wheatley, J.~M., Welsh, B.~Y., \& Browne, S.~E.\ 2008, \aj, 136, 259. doi:10.1088/0004-6256/136/1/259
\bibitem[Wils et al.(2010)]{Wils_novae_2010} Wils, P., G{\"a}nsicke, B.~T., Drake, A.~J., et al.\ 2010, \mnras, 402, 436. doi:10.1111/j.1365-2966.2009.15894.x
\bibitem[Woodgate et al.(1998)]{stis_woodgate_1998} Woodgate, B.~E., Kimble, R.~A., Bowers, C.~W., et al.\ 1998, ''The Space Telescope Imaging Spectrograph Design", \pasp, 110, 1183 

\end{thebibliography}
\end{document}